\documentclass[twocolumn,preprintnumbers,amsmath,amsfonts,amssymb]{revtex4}
\usepackage[pdfstartview=FitH,linkcolor=blue,citecolor=red,urlcolor=green]{hyperref}
\usepackage{graphics,epsfig,graphicx}% Include figure files
\usepackage{dcolumn}% Align table columns on decimal point
\usepackage{bm}% bold math
\usepackage{color}

\newcommand{\Ignore}[1]{}
\usepackage{epstopdf}
\begin{document}

\title{Magnetic decoupling of ferromagnetic metals through a graphene spacer}

\author{I. Grimaldi$^{1}$}
\author{M. Papagno$^{1}$}
\author{L. Ferrari$^{2, 3}$}
\author{P. M. Sheverdyaeva$^{3}$}
\author{S. K. Mahatha$^{3}$}
\author{D. Pacil\'{e}$^{1, 3}$\footnote{E-mail address: daniela.pacile@fis.unical.it}}
\author{C. Carbone$^{3}$}

\affiliation{%
$^{1}$~\mbox{Dipartimento di Fisica, Universit$\grave{a}$ della Calabria, 87036 Arcavacata di Rende (CS), Italy} \\
$^{2}$~\mbox{Istituto dei Sistemi Complessi, Consiglio Nazionale delle Ricerche, I-00133 Roma, Italy}\\
$^{3}$~\mbox{Istituto di Struttura della Materia, Consiglio Nazionale delle Ricerche, Trieste, Italy}\\
}%

%\date{\today}
\begin{abstract}
\noindent
We study the magnetic coupling between different ferromagnetic metals (FMs) across a graphene (G) layer, and the role of graphene as a thin covalent spacer. Starting with G grown on a FM substrate (Ni or Co), we deposit on top at room temperature different FM metals (Fe, Ni, Co). By measuring the dichroic effect of 3\textit{p} photoemission lines we detect the  magnetization of the substrate and the sign of the exchange coupling in FM overlayer at room temperature. We show that the G layer magnetically decouples the FM metals. 
\end{abstract}

\maketitle

%\newpage
\section{Introduction}
The electronic transport and the magnetic coupling between different magnetic species separated by a graphene layer (G) has recently triggered interest for spintonics applications \cite{ Cobas, Klar, Candini, Vo-Van, Lisi}. Spin-dependent transport perpendicular to the graphene plane results in tunneling magnetoresistance.  
Concerning the behavior of different ferromagnetic metals (FMs) chemically separated by a G layer, it has been predicted that the coupling can be either ferromagnetic or antiferromagnetic, depending on the choice of the two species \cite{Yazyev}. This behavior might provide the opportunity for tuning the coupling by varying the chemical composition of the FM/G/FM junction. 

By recent X-ray magnetic circular dichroism (XMCD) measurements \cite{Barla} on FM single atoms or islands deposited on G/Ni at T$<$5K, we observed a complex behavior of G in transmitting the exchange interactions through the perpendicular direction. Combined experimental and theoretical results have shown that the sign of the magnetic coupling depends on the adsorption sites for isolated atoms and varies in a complex way for dimers and larger aggregates. Accordingly, transitions between anti- and ferromagnetic coupling, depending on the geometrical arrangement of atoms at surface, have been observed. Those results prove a high sensitivity of the G-mediated exchange interaction to the local coordination of the adatoms.

Here, we examine the magnetic coupling between FM metals (Fe, Co, Ni), deposited at room temperature (RT) and measured in remanent magnetization conditions, on G/Co(0001) and G/Ni(111) substrates. We employ the linear dichroic effect in 3\textit{p} photoemission lines to examine the magnetic coupling between FMs across a G layer. By measuring the line shape of 3\textit{p} core levels as a function of the direction of magnetization and along different symmetry directions of the substrate, we are able to detect the  magnetic coupling across the graphene layer or its absence. Our results show that for the selected FM/G/FM junctions, grown and measured at RT, no magnetic coupling is attained  through the G layer. Therefore, the G inhibitis the magnetic alignment that normally occurs when two FMs are put in contact.

\begin{figure*}
\scalebox{0.65}{\includegraphics{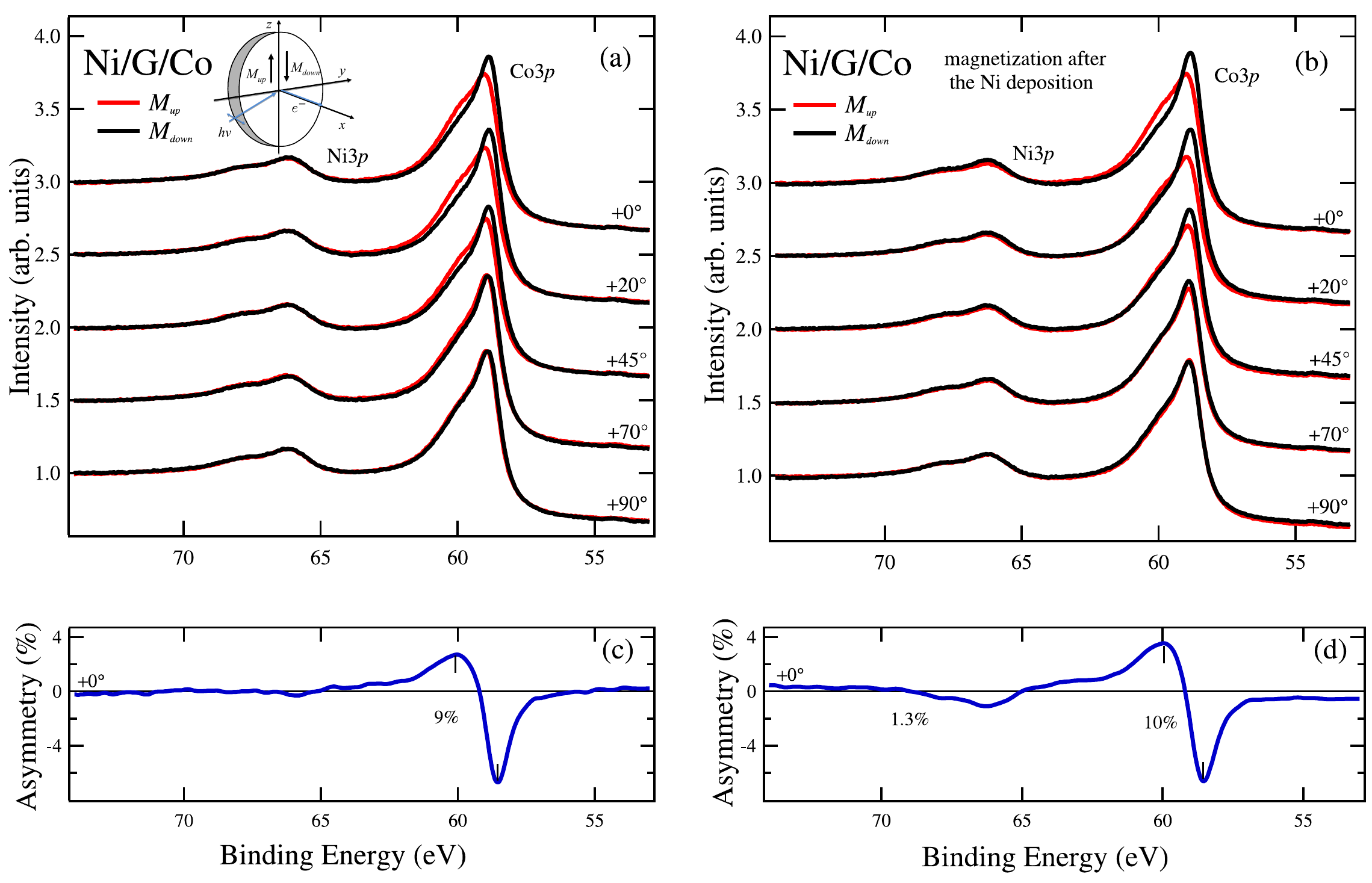}}
\vskip -10pt
\caption{\label{fig2}
(Color online) Core level measurements taken on (a) 0.2 ML Ni/G/Co in remanent magnetization prior to the Ni deposition, and (b) after the Ni deposition, showing
Co 3\textit{p} and Ni 3\textit{p} photoemission peaks acquired along selected azimuthal directions and for different magnetizations (\textit{M up} and \textit{M down}); (c-d) asymmetry curves extracted from data taken at +0$^{\circ}$ of panels (a) and (b), respectively. In the inset of Fig.1a the MLD chiral geometry is reported.}
\end{figure*}

\section{Methods}\label{sec2:methods}

The experiments were performed at the VUV beamline of the Elettra synchrotron radiation laboratory~(Trieste, Italy) using a Scienta R4000 electron energy analyzer at a base pressure of $5\times 10^{-11}$\,mbar. 
The W(110) crystal was prepared by repeated high temperature flash-annealing cycles in oxygen atmosphere until a sharp 1$\times$1 low-energy electron diffraction (LEED) pattern and well-developed W 4\textit{f} surface core level components were observed. The Co(0001) and Ni(111) films were grown on W(110) by evaporation of 10$\div$25 ML of Co (or Ni) at RT from an electron-bombarded rod. The LEED showed a 1$\times$1 hexagonal pattern for both Co and Ni films, with sharp diffraction spots upon annealing at about 420 K. 
The graphene layer was grown by CVD of ethylene at about 700 K, at the lower limit of temperature to avoid local breaking of the film \cite{Pacile}. Surface order and cleanliness were checked by LEED and core level measurements. The samples were magnetically saturated by applying a pulsed current to a coil wrapped around the sample holder, and always measured in remanence at RT.
FMs (Fe, Co or Ni) were deposited on G/Co (or G/Ni) at RT. In this experimental conditions FMs tend to form 3D clusters without long-range order \cite{Lahiri, Sicot, Liu, Sarkar, Liu2}. Starting from the submonolayer regime, the FM coverage was here estimate on the bare substrate by measuring the intensity ratio of  3\textit{p} photoemission lines. Core levels photoemission spectra were collected using a photon energy of 150 eV. The scattering geometry of our experiments is reported as inset of Fig. 1a. Light  with linear horizontal polarization was impinging the sample at 45$^{\circ}$ from the surface normal. Electrons were collected under normal emission with about 30$^{\circ}$ full acceptance along the scattering plane (\textit{xy}). The Co or Ni easy magnetization axis is set to lie at different angles with respect to the scattering plane by means of azimuthal rotation of the sample. If the easy magnetization axis is oriented along \textit{z}, namely perpendicularly to the scattering plane, core levels spectra exhibit a large dichroic signal when the magnetization is reversed from positive (\textit{M up}) to negative (\textit{M down}) \textit{z} values. The Ni(111) and Co(0001) films were found to be in-plane magnetized along the [1$\bar{1}$0] and [1$\bar{1}$00] symmetry directions, respectively. However, when the magnetic pulse was not applied in proximity of the easy axis, the largest dichroic signal was collected away from the perpendicular direction. For clearity, in these cases symmetry directions were not reported. 
In a typical Magnetic Linear Dichroism (MLD) experiment, the magnetization direction is simply reversed (e.g. by reversing the current through the coil) and this causes to first order to a  change of the relative intensities of multiplet sublevels, due to dipole matrix element effects \cite{G. van der Laan2}. The dichroic signal is obtained by calculating the asymmetry, i.e. the difference between spectra acquired with different magnetization divided by their sum, A=(I$^{\textit{M up}}$-I$^{\textit{M down}}$)/(I$^{\textit{M up}}$+I$^{\textit{M down}}$). This quantity is proportional to the surface magnetization, i. e. to the average orientation of magnetic moments at surface, thus defining an element-specific order parameter.

\begin{figure}
\scalebox{0.65}{\includegraphics{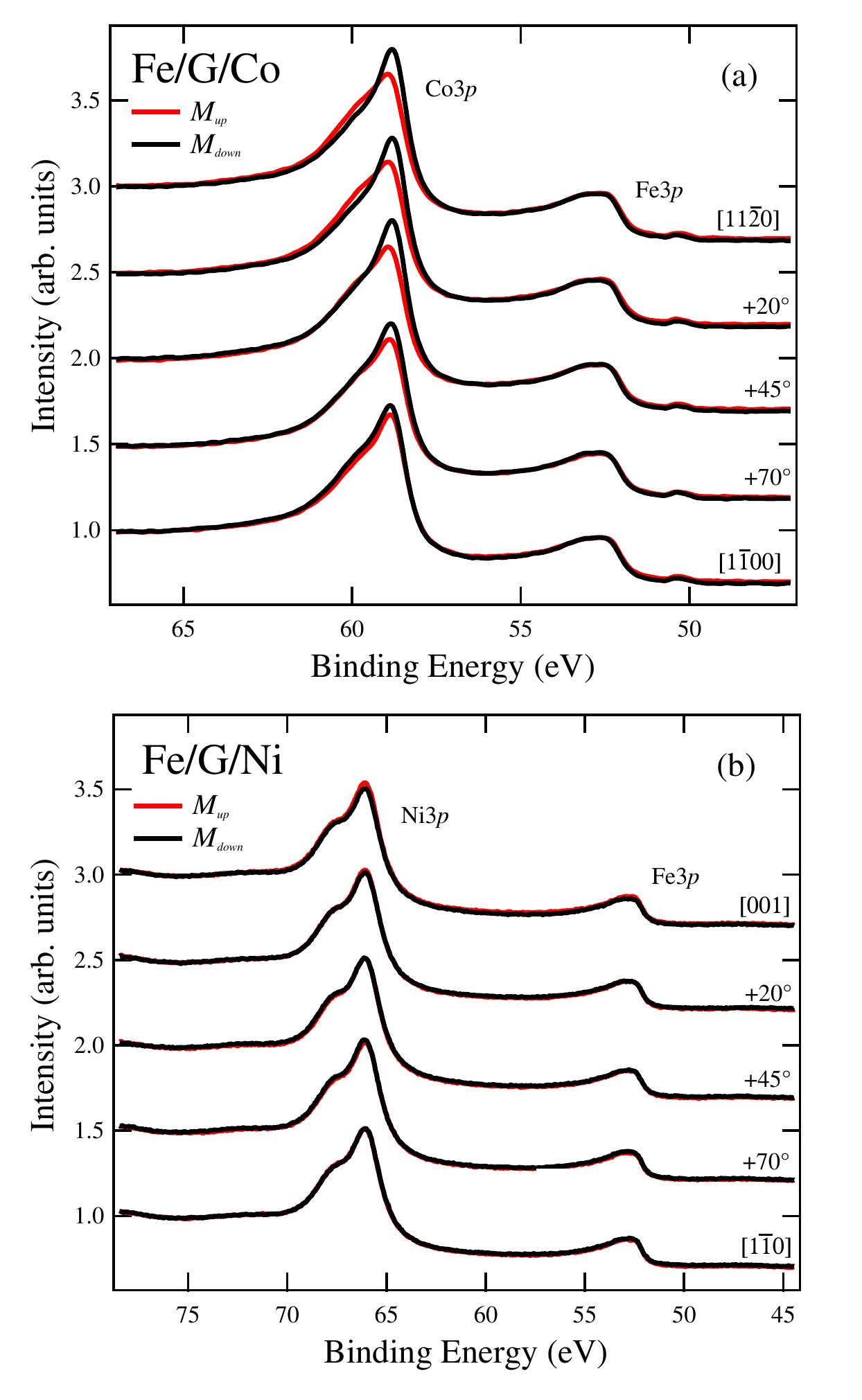}}
\vskip -10pt
\caption{\label{fig3}
(Color online)
Core level measurements taken on (a) 0.4 ML Fe/G/Co, (b) 0.3 ML Fe/G/Ni, showing 3\textit{p} photoemission peaks acquired along selected azimuthal directions and for different magnetizations (\textit{M up} and \textit{M down}).}
\end{figure}

\begin{figure}
\scalebox{0.65}{\includegraphics{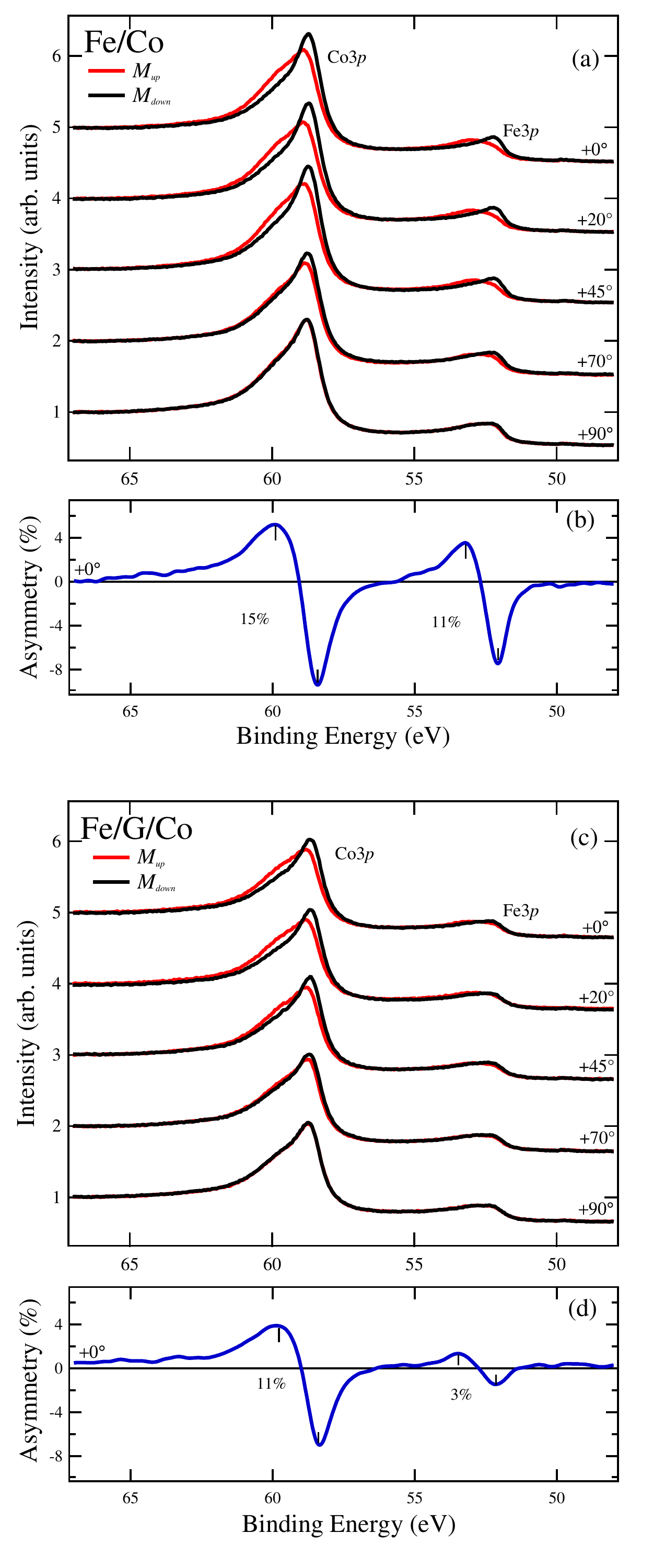}}
\vskip -10pt
\caption{\label{fig4}
(Color online)
Core level measurements taken on (a) 0.2 ML Fe/Co, (c) partially covered 0.2 ML Fe/0.9 ML G/Co, showing Fe3\textit{p} and Co3\textit{p} photoemission peaks acquired along selected azimuthal directions and for different magnetizations; (b-d) asymmetry curves extracted from data taken +0$^{\circ}$ of panels (a) and (c), respectively.}
\end{figure}

\section{Results and Discussion}
In Fig. 1, MLD measurements taken on the Ni/G/Co system are reported. The sample was prepared by depositing 0.2 ML of Ni on the G/Co substrate remanently magnetized (Fig. 1a). The line shape of Co 3\textit{p} and Ni 3\textit{p} photoemission peaks is monitored along different azimuthal directions, and pair spectra were obtained by reversing the magnetization. The dichroic signal is present for the Co 3\textit{p} photoemission peak, while is not observed for the Ni 3\textit{p} peak, as confirmed by the asymmetry curve reported in Fig. 1c. The dichroism measured at the Co 3\textit{p} peak consists in a plus/minus feature about 1.5 eV wide, and a maximum peak-to-peak asymmetry of about 9\%, in agreement with previous results \cite{Gomoyunova}. Extrema (minimum and maximum) correspond to sublevels of the Co 3\textit{p} multiplet \cite{GvanderLaan3}. Our MLD results show that while the Co substrate is magnetized, no magnetic coupling is transmitted to the Ni clusters. The capability of the clusters to align their magnetic moments was checked by applying a magnetic pulse to the Ni/G/Co system and measuring the core levels (Fig. 1b). As shown in Fig. 1d, a detectable asymmetry of Ni 3\textit{p}  is extracted (1.3\%) and the similarity between dichroic signals, with a prominent minus feature, indicates that Ni and Co  moments are parallelly aligned. 

Similar experiments were performed with Fe clusters deposited on top of G/Co (Fig. 2a) and G/Ni (Fig. 2b) substrates. With respect to the Ni 3\textit{p} photoemission peak, the Fe 3\textit{p} one exhibits a strong dichroic behavior \cite{Roth}, allowing to detect also faint line shape variation induced by the substrate. Similarly to the experiment of Fig. 1a, we show in Fig. 2 that while the Co and Ni substrates exhibit the expected magnetization, with the easy magnetization axis along [1$\overline{1}$00] and [1$\overline{1}$0], respectively, no dichroic effect is seen for the Fe 3\textit{p} photoemission peak. 

In order to investigate further the effect of graphene in FM/G/FM junctions, we compare previous results with those obtained for a partially covered FM substrate. As a starting point, we show in Fig. 3a the dichroic behavior of Fe clusters (0.2 ML) deposited on a pristine Co film. As expected, Co atoms and Fe overlayer produce a significant ferromagnetic coupling (Fig. 3b). Further, we show a partially covered G/Co surface (Fig. 3c), achieved by means of exposures to ethylene below the saturation condition. 
The uncovered Co surface was estimated to be below 10\% by evaluating the ratio of C 1s/Co 3p photoemission peaks, taken at about 700 eV of photon energy (data not shown), with respect to the ratio obtained at the saturation coverage. Comparing corresponding MLD measurements with those obtained for a fully covered substrate (Fig. 2a), it is deduced that the Fe-Co ferromagnetic coupling seen in Fig. 3d is due to a direct contact of the two metals in the reduced uncovered regions. Therefore, we conclude that the presence of G in our junctions weakens the exchange interactions between FMs, leading to a decoupling of their magnetic moments. 

The electronic and magnetic properties of G/Ni(111) and G/Co(0001) have been widely characterized \cite{Usachov, Marchenko}. The Dirac cone of G is split in several parts above and below the Fermi level due to the interaction between C 2\textit{p$_{z}$} and \textit{d} state of the metallic substrate. These interface states with a notably contribution of graphene \textit{$\pi$}-orbitals acquire a partial spin polarization by the magnetic substrates, due to hybridization with distinct majority and minority Co (or Ni) \textit{d} bands. As consequence, in a defect-free G/Ni system the net magnetic moment of carbon was evaluated to be of about 0.01 $\mu_{B}$ per atom \cite{Bhandary, Marocchi}, taking into account that in the two different absorption sites on Ni(111) carbon atoms acquire opposite polarization compared to the substrate. The magnetic interactions through graphene is based on direct exchange, and has therefore a different origin from the well-known interlayer oscillatory coupling through non-magnetic and metallic spacers. In a previous XMCD experiments \cite{Barla}, performed on Co/G/Ni junctions, grown and measured at T$<$5 K, we observed a significant G-mediated exchange coupling (in the range of tens of meV for Co atom) between Co atoms and Ni(111).
In the present experiments, we found that such G-mediated exchange coupling is not sufficient to align the magnetic moment of the deposited metal at RT. On one side, the magnetic coupling change sign and is mediated between different adsorption sites. The effective magnitude of the coupling is furtherly reduced for RT grown overlayers where 3D clustering may be favoured with respect to a 2D growth. If cluster growth prevails, the number of atoms in contact with graphene is reduced and the effective exchange interaction proportionally decreases. The absence of adlayer magnetization suggests that disordered ferromagnetic domains are formed but point in different spatial directions or, less likely, the presence of fluctuating paramagnetic units. In fact, by applying a magnetic pulse to the Ni/G/Co junction (Fig. 1b), we found that the applied field partially  orient the  magnetization of the overlayer to the substrate, pointing out that at least a fraction of the metal is  blocked with respect to thermal fluctuations. 

The here reported decoupling of the magnetic elements suggest that further developments in the realization of nanoscale magnetic structures based on tunneling magneto resistance through graphene may be achieved.  A practicable objective appears to be the production of patterned ferromagnetic elements of nanometric size, separate from a magnetically hard material by a graphene layer, and acting as independent tunneling magneto-resistance units.

%\section*{Acknowledgements}
%This work ........

\end{document}